\documentclass[journal,10pt,twocolumn]{IEEEtran}
\usepackage{graphicx}
\usepackage{epsfig}
\usepackage{subfigure}
\usepackage{amssymb}
\usepackage{amsbsy}
\usepackage{amsmath}
\usepackage{cite}
\usepackage{url}

\usepackage{color}
\usepackage{float}
\usepackage{times,amsmath,epsfig}
\usepackage{xspace,latexsym,syntonly}
\usepackage{amssymb}
\usepackage{amsfonts}
\usepackage{textcomp}
\usepackage{subfigure}
\usepackage{amsbsy}
\usepackage{cite}

\newtheorem{theorem}{Theorem}
\newtheorem{lemma}{Lemma}
\newtheorem{remark}{Remark}
\newtheorem{definition}{Definition}

\newcommand{\beq}{\begin{equation}}
\newcommand{\eeq}{\end{equation}}
\newcommand{\bea}{\begin{array}}
\newcommand{\ena}{\end{array}}
\newcommand{\bds}{\begin {itemize}}
\newcommand{\eds}{\end {itemize}}
\newcommand{\bdf}{\begin{definition}}
\newcommand{\blm}{\begin{lemma}}
\newcommand{\edf}{\end{definition}}
\newcommand{\elm}{\end{lemma}}
\newcommand{\bthm}{\begin{theorem}}
\newcommand{\ethm}{\end{theorem}}
\newcommand{\bprp}{\begin{prop}}
\newcommand{\eprp}{\end{prop}}
\newcommand{\bcl}{\begin{claim}}
\newcommand{\ecl}{\end{claim}}
\newcommand{\bcr}{\begin{coro}}
\newcommand{\ecr}{\end{coro}}
\newcommand{\bquest}{\begin{question}}
\newcommand{\equest}{\end{question}}

\newcommand{\larrow}{{\larrow}}

\newcommand{\nin}{{\not \in}}

\def\urltilda{\kern -.15em\lower .7ex\hbox{\~{}}\kern .04em}

\begin{document}

\title{Spectrum and Energy Efficient Multiple Access for Detection in Wireless Sensor Networks}

\author{Kobi Cohen, Amir Leshem
\thanks{This work has been submitted to the IEEE for possible publication. Copyright may be transferred without notice, after which
this version may no longer be accessible.}
\thanks{A short version of this paper was presented at IEEE International Symposium on Information Theory (ISIT) 2018 \cite{cohen2018density}.}
\thanks{Kobi Cohen is with the Department of Electrical and Computer Engineering, Ben-Gurion University of the Negev, Beer-Sheva 84105, Israel. Email: yakovsec@bgu.ac.il}
\thanks{Amir Leshem is with the Faculty of Engineering, Bar-Ilan University, Ramat Gan 52900, Israel. Email: leshem.amir2@gmail.com}
\thanks{The work of K. Cohen was supported by the U.S.-Israel Binational Science Foundation (BSF) under grant 2017723. The work of A. Leshem was supported by the ISF grant 903/2013.}
}
\date{}
\maketitle

\begin{abstract}

We consider a binary hypothesis testing problem using Wireless Sensor Networks (WSNs). The decision is made by a fusion center and is based on received data from the sensors. We focus on a spectrum and energy efficient transmission scheme used to reduce the spectrum usage and energy consumption during the detection task. We propose a Spectrum and Energy Efficient Multiple Access (SEEMA) transmission protocol that performs a censoring-type transmission based on the density of observations using multiple access channels (MAC). Specifically, in SEEMA, only sensors with highly informative observations transmit their data in each data collection. The sensors transmit a common shaping waveform and the fusion center receives a superposition of the analog transmitted signals. SEEMA has important advantages for detection tasks in WSNs. First, it is highly energy and bandwidth efficient due to transmission savings and narrowband transmission over MAC. Second, it can be implemented by simple dumb sensors (oblivious to observation statistics, and local data processing is not required) which simplifies the implementation as compared to existing MAC transmission schemes for detection in WSNs. We establish a finite sample analysis and an asymptotic analysis of the error probability with respect to the network size and provide system design conditions to obtain the exponential decay of the error. Specific performance analysis is developed for common non-i.i.d. observation scenarios, including local i.i.d. observations, and Markovian correlated observations. Numerical examples demonstrate SEEMA performance.
\end{abstract}

\section{Introduction}
\label{sec:introduction}

We consider a binary detection problem in WSNs in which sensor nodes measure a certain phenomenon and upon request (i.e., a data collection event) transmit some function of their observations to the fusion center (FC) through a block fading channel. The FC makes decisions whether an unknown hypothesis is $H_0$ or $H_1$ based on the received data. We assume that observation statistics is only available at the FC\footnote{Learning the observation statistics is typically done by scattering reference nodes in the field \cite{Hong_Channel_2008}.}. The sensor nodes can be simple and dumb \cite{Marano_DOA_2005, Liu_Type_2007} and are not aware of their task or the environment characteristics.

\subsection{Main Results}

\noindent
\textbf{Algorithm Development:} We propose a Spectrum and Energy Efficient Multiple Access (SEEMA) transmission protocol that performs a censoring-type transmission scheme based on the density of observations using multiple access channels (MAC). Specifically, in SEEMA, only sensors with highly informative observations transmit their data in each data collection. The sensors transmit a common shaping waveform and the fusion center receives a superposition of the analog transmitted signals. We propose a closed-form threshold-based detector that requires observation statistics only at the FC (and not at each sensor as required by the Likelihood-Based Multiple Access (LBMA) scheme described in Section \ref{ssec:related}).

\noindent
\textbf{Efficient and Low-Complexity Implementation:} SEEMA has important advantages for detection tasks in WSNs. In practical implementations of WSN tasks reducing the number of transmitted sensors is a key goal for reducing the energy consumption involved in each data collection. Thus, SEEMA performs censoring-type transmissions which lead to significant energy saving in this respect. In the traditional communication approach to detection in WSNs, sensors transmit some function of their observations over parallel channels (for instance, FDM/TDM fashion). However, the bandwidth increases linearly with the number of sensors in this scheme. Therefore, for a large-scale WSN, transmission over multiple access channels (MAC) is advantageous. By using MAC in SEEMA, all sensors transmit simultaneously in one dimension (or a small number of dimensions). As a result, the bandwidth requirement does not depend on the number of sensors. Implementing the threshold-based detector is simple and does not require computing a complex rate function that depends on the channel distribution of each sensor node as is the case for the Type Based Multiple Access (TBMA) scheme \cite{Mergen_Asymptotic_2007} described in Section \ref{ssec:related}. Furthermore, the bandwidth usage does not depend on data dimension size, unlike the TBMA scheme (that uses MAC as well), where the bandwidth increases linearly with the number of (independent) data dimensions. Finally, SEEMA can be implemented by simple dumb sensors (oblivious to observation statistics and local data processing is not required) which simplifies the implementation as compared to existing MAC transmission schemes for detection in WSNs, e.g., TBMA and LBMA (a detailed discussion of existing methods appears in Section \ref{ssec:related}).

\noindent
\textbf{Performance Analysis:} We establish both finite sample analysis and asymptotic analysis of the error probability with respect to the network size and provide system design conditions for obtaining exponential decay of the error. Our analysis is valid for models with additive white sub-Gaussian noise, which is more general than the classic AWGN model. Specifically, we use large deviation (LD) theory to characterize the detector's error exponent when the number of sensor nodes approaches infinity. We also establish performance bounds on the error probability for a finite number of sensor nodes. For the case of i.i.d. observations and equal channel gains, we provide tighter finite-sample bounds that coincide with the asymptotic error exponent. By contrast, under TBMA, there is a gap between the finite sample bounds and the asymptotic error exponent \cite{Mergen_Asymptotic_2007, Liu_Type_2007}. Specific performance analysis is developed for common non-i.i.d. observation scenarios, including local i.i.d. observations, and Markovian correlated observations. Numerical experiments then demonstrate SEEMA performance.

\subsection{Related Work}
\label{ssec:related}

Event detection has attracted much attention in the field of WSNs in past and recent years. Available methods and technology appear in \cite{Akyildiz_A_survey_2002, sujithra2017survey, puzanov2018deep} and references therein. Developing energy and spectrum efficient transmission protocols for WSNs has attracted much attention in past and recent years. In traditional communication protocols for inference tasks in WSNs, each sensor transmits using orthogonal channels (e.g., FDM/TDM). Such methods have focused on various ways to reduce spectrum and energy consumption. In \cite{tsitsiklis1986threshold}, the focus was on sensors that measure conditionally i.i.d. observations and transmit a binary function of their observations (based on the likelihood-ratio information) to a fusion center (through parallel channels with equal gains) which then decides which one of two alternative hypotheses is true. Refinements and asymptotic analysis of the detection error have been established in \cite{tsitsiklis1989decentralized}. In this paper, however, the focus is on transmissions through multiple access fading channels, the observation distributions are assumed to be known only at the FC, observations can be non-i.i.d., and both finite and asymptotic analysis are derived. In \cite{Zhao_Opportunistic_2005, Chen_An_2007, Cohen_TOP_ICASSP_2009, Cohen_TOP_2010}, the focus was on exploiting the channel diversity among sensors by scheduling sensors that experienced better channels for transmission to reduce the transmission energy. Active fusion strategies for event detection have been developed in \cite{cohen2015active, cohen2015asymptotically, huang2018active}. In \cite{Rago_Censoring_1996, Appadwedula_Decentralized_2007, Patwari_Hierarchical_2003}, measures of the quality of observations for scheduling sensors with better informative observations were exploited to reduce the number of transmissions. This approach is also known as \emph{censoring} \cite{Rago_Censoring_1996}. A distributed access protocol that reduces the number of transmissions by ordering transmissions according to the magnitude of the log likelihood ratio was proposed in \cite{Blum_Energy_2008, zhang2017ordering}. In our previous work we developed a method that combines both channel state and quality of observations to achieve energy savings \cite{Cohen_Likelihood_2010, Cohen_Energy_2011}. In \cite{braca2011asymptotically, braca2012single}, the authors proposed a detection scheme that only uses one transmission based on the highest magnitude of the log likelihood ratio, and showed that it is asymptotically consistent. However, these schemes require knowing the observation statistics at the sensor nodes for local data processing, which is assumed to be known only at the FC in this paper. Furthermore, the bandwidth increases linearly with the network size when using schemes that transmit on parallel channels (i.e., dimension per sensor). Therefore, for large-scale WSNs, transmissions over multiple access channels (MAC) is advantageous in terms of bandwidth efficiency, which is why this is the focus of this paper. In \cite{niu2006distributed}, the authors investigated a counting rule that counts local binary decisions of a DC signal in noise model. In \cite{sundaresan2011copula}, copula-based fusion was investigated for detection under correlated observations. However, it requires transmissions over parallel channels and the complexity increases exponentially with the network size. Low-complexity approximations were proposed in \cite{sundaresan2011copula}. In \cite{chen2004channel, chen2006channel}, channel-aware methods for detection were investigated.

It is well known that digital communication (where sensor nodes convert their observations into a bit stream) does not lead to optimal performance in general network problems. The correct way of understanding the nature of information is in an analog form, rather than as bits \cite{gastpar2008uncoded}. In \cite{nazer2007computation}, joint source-channel strategies over MAC were developed that often outperformed separation-based strategies. A well known transmission scheme that uses MAC for detection is Likelihood Based Multiple Access (LBMA) \cite{Liu_Type_2007, cohen2013performance} (which was also used for estimation tasks in \cite{Marano_Likelihood_2007}). In LBMA, each sensor computes the log-likelihood ratio (LLR) locally based on its current random observation, and then amplifies the transmitted waveform by the LLR. However, computing the LLR locally requires knowing the distribution observation under each hypothesis at each sensor, which is assumed to be known only at the FC in this paper. Furthermore, the hardware implementation is more complex than SEEMA since transmitting the random LLRs, which have a large dynamic range, can cause signal distortion due to a saturation effect in the analog amplifiers. By contrast, in SEEMA the transmitted waveform amplitude is deterministic, which is a desired property in analog transmissions. A well-known access scheme that can be implemented by dumb sensors is termed Type Based Multiple Access (TBMA) \cite{Mergen_Asymptotic_2007, Liu_Type_2007}. In TBMA, the observations are quantized before communication to $K$ possible levels. Sensors that observe level $k$ transmit a corresponding waveform $k$ from a set of $K$ orthonormal waveforms. In each data collection all the sensors transmit their waveforms in a one-shot transmission and the FC receives a superposition of the waveforms over MAC. In the TBMA scheme, observation statistics is only needed at the FC. In terms of bandwidth requirement, TBMA is much less efficient than SEEMA. The bandwidth requirement grows linearly with $K$ and the number of (independent) data dimensions $d$ (since each dimension must be quantized and transmitted to obtain its type at the FC). By contrast, under SEEMA, the bandwidth requirement is independent of $d$. In terms of the number of transmissions, under TBMA, all sensors participate in each data collection, whereas SEEMA performs censoring-type transmissions. Generalizations of TBMA using non-coherent transmissions and i.i.d. observations were studied in \cite{anandkumar2007type, li2011decision}. However, here we assume coherent transmissions by phase correction at the transmitter as in \cite{Mergen_Asymptotic_2007, cohen2013performance, wimalajeewa2015wireless} and the non-i.i.d. observation case, which make the problem fundamentally different. Other related works have investigated MAC for detection in WSN using multiple antennas at the FC \cite{nevat2014distributed}, detection with a non-linear sensing behavior \cite{zhang2016event}, and detecting a stationary random process distributed in space and time with a circularly-symmetric complex Gaussian distribution \cite{maya2015optimal, maya2015exploiting}. However, these studies are fundamentally different from the settings considered in this paper.

\subsection{Organization}
The remainder of this paper is organized as follows. In section \ref{sec:network} we present the network model, and present the SEEMA scheme and the proposed detector. In section \ref{sec:performance} we detail the theoretical performance analysis of the algorithm. In section \ref{sec:simulations} we provide simulation results.

\section{Detection Scheme using Spectrum and Energy Efficient Multiple Access (SEEMA)}
\label{sec:network}

We consider a binary detection problem using a WSN containing $N$ sensors. The sensors measure a certain phenomenon and deliver some function of their observations to a FC through a multiple access channel. We assume that sensor $n$ experiences a block fading channel $h_n$ with a non-zero channel mean\footnote{As explained in the introduction, this is done by correcting the phase at the transmitter.} $\mu_{h,n}$. The FC determines whether an unknown hypothesis is $H_0$ or $H_1$ based on the received data from the sensors. The a-priori probabilities of the two hypotheses $H_0$, $H_1$ are denoted by $P(H_0)$ and $P(H_1)$, respectively. Let $x_n$ and $f_{X_n}( x | H_m )$ be the random observation (vector) at sensor $n$ and the Probability Density Function (PDF) of $x_n$ conditioned on $H_m$, respectively.

\subsection{Transmission Scheme}

Under SEEMA, all sensors that observe $x_n$ in a predetermined transmission region of observations transmit a common waveform. Let $\Gamma_n$ be the (multi-dimensional) transmission region of sensor $n$ observation, and let
\beq
\label{eq:p_0_n_p_1_n}
\bea{l}
\vspace{0.0cm}
p_{0,n}\triangleq\displaystyle\int_{x\in\Gamma_n}{f_{X_n}( x | H_0 )}dx, \vspace{0.2cm} \\
p_{1,n}\triangleq\displaystyle\int_{x\in\Gamma_n}{f_{X_n}( x | H_1 )}dx \;,
\ena
\eeq
so that $p_{i,n}$ is the probability that sensor $n$ transmits under $H_i$. We design $\Gamma_n$ such that $p_{1,n}>p_{0,n}$ for all $n$ (i.e., it is more likely to transmit when event $H_1$ occurs).
In practice, the transmission region is predetermined by the FC based on the density of observations to increase the distance between the hypotheses under constraint on the expected number of transmissions. A discussion about design principles of $\Gamma_n$ is given later. Throughout the paper we will focus on detector performance assuming that $\Gamma_n$ is given.

Let $s(t), \;0<t<T$ be a baseband equivalent normalized waveform, $\int_{0}^{T}{s^2(t)dt}=1$. In each data collection, all sensors that observe $x_n$ in the transmission region $\Gamma_n$ transmit $A_n\sqrt{E_N}\cdot s(t)$. None of the other sensors transmit. $E_N$ can be any fixed constant or a function of the number of sensors $N$, such that the power constraint is satisfied. $A_n$ is a finite amplification and is given by:
\beq
\bea{l}\label{eq:A_n}
    A_n =\log\left(\frac{(1-p_{0,n})p_{1,n}}{(1-p_{1,n})p_{0,n}}\right)e^{-j\phi_h}\;.
\ena
\eeq
where $e^{-j\phi_h}$ is due to phase correction at the receiver as in \cite{Mergen_Asymptotic_2007, cohen2013performance, wimalajeewa2015wireless}.
The motivation for amplifying the signal by $A_n$ is to enable SEEMA to achieve the best error exponent which is obtained by the maximum likelihood detector with respect to the transmitted signal when the observations are independent and the channel gains are equal across sensors, as shown in Theorem \ref{th:no_fading}.c. It should be noted that phase correction is only needed to produce channel gains with nonzero means at the receiver. In the case where the channel gains have nonzero means, $\phi_h$ can be set to zero and phase correction is not required\footnote{The exact expressions for the error exponent in the analysis holds under the ideal assumption that the channel phase is completely corrected. However, receiving non-zero mean signals is sufficient to achieve the same order of decay (i.e., exponential decay of the error probability for $E_n=\Omega(N^{-1})$ and sub-exponential decay for $E_n=\Omega(N^{\epsilon-2})$, for any fixed $\epsilon>0$).}.

In the case where the channel gains have zero mean, correcting the phase with an error less than $\pi/4$ is sufficient to yield channel gains with nonzero means at the receiver. Therefore, only partial information about the channel phase is required. Let $\textbf{1}_{\Gamma_n}(x_n)=1$ if $x_n\in\Gamma_n$, or $\textbf{1}_{\Gamma_n}(x_n)=0$ if $x_n\nin\Gamma_n$ be the indicator function. The received signal at the FC is given by:
\beq
\label{eq:r_t}
\bea{l}
    r(t) =\displaystyle\sum_{n=1}^{N}{h_n A_n\textbf{1}_{\Gamma_n}(x_n) \sqrt{E_N}\cdot s(t) }+w(t) \;\;, 0<t<T\;,
\ena
\eeq
where $w(t)$ is a zero-mean additive interference, and $h_n$ is a non-zero mean r.v. due to phase correction. \\
After matched-filtering by the corresponding waveform at the FC, we have:
\beq
\label{eq:r}
\bea{l}
r=\sqrt{E_N}\displaystyle\sum_{n=1}^{N}{h_n A_n \textbf{1}_{\Gamma_n}(x_n)}+w,
\ena
\eeq
where $w\sim subG(\sigma^2)$ is a zero-mean $\sigma^2$-sub-Gaussian r.v. (see Remark \ref{remark:subG} for more details). Let
\beq
\label{eq:y}
\bea{l}
y_{_N}\triangleq\displaystyle\frac{r}{N\sqrt{E_N}}=\displaystyle\frac{1}{N}\sum_{n=1}^{N}{h_n A_n \textbf{1}_{\Gamma_n}(x_n)}+\tilde{w},
\ena
\eeq
where $\tilde{w}\sim subG(\sigma^2/N^2E_N)$. \\

We propose the following threshold-based detector:\\
\textit{Decide $H_1$ if:}
\beq
\label{eq:detector}
\bea{l}
\displaystyle\frac{y_{_N}}{Z}>\displaystyle\frac{\log(\eta)}{N}+\frac{1}{N}\sum_{n=1}^{N}{\log\left(\frac{1-p_{0,n}}{1-p_{1,n}}\right)}
 \;.
\ena
\eeq
\textit{Otherwise, decide $H_0$}.\\
We use a threshold-based detector since it is practically appealing, achieves the desired decay of the error probability, and maximizes the error exponent as the network size increases and the channels have equal gains, as shown in the analysis in Section \ref{sec:performance}.
The term $Z>0$ is a normalization constant and is discussed in Section \ref{sec:performance}. Under the MAP criterion $\eta=P(H_0)/P(H_1)$, and under the Neyman Pearson (NP) criterion $\eta$ is determined according to the desired false-alarm probability.

\begin{remark}
\label{remark:subG}
Note that our model takes into account a sub-Gaussian interference model. Specifically, a random variable $w$ is said to be $\sigma^2$-sub-Gaussian with variance proxy $\sigma^2$ if it has zero mean and its moment generating function (MGF) satisfies $E[e^{sw}]\leq e^{\sigma^2s^2/2}$ for all $s\in\mathbb{R}$. Clearly, the classic AWGN model is a special case of our model, in which $w\sim\mathcal{N}(0, \sigma^2)$. A more general common communication model that our model captures is the case where the desired signal is attenuated by the fading channel and received with an additive external bounded interferer plus white Gaussian noise \cite{heimann2016non}.
\end{remark}

\subsection{Implementation of SEEMA}
\label{ssec:implementation}

The implementation of the SEEMA scheme has important advantages for detection using WSN. It is highly bandwidth-efficient because only a single waveform $s(t)$ is transmitted by the sensors. Furthermore, the number of transmissions can be significantly reduced depending on the desired detection performance (see Section \ref{sec:performance}) and system constraints. In practical implementations of WSN tasks reducing the number of transmitted sensors is a key goal for reducing the energy consumption involved in each data collection. Unlike LBMA and TDMA that use all sensors for transmitting data in each data collection, SEEMA applies self censoring-type transmissions over MAC, where only sensors that measure observations which lie inside the transmission region participate in the data collection. Therefore, SEEMA is an energy-efficient scheme in this respect.

We point out that SEEMA readily applies to the case where sensors measure $d$-dimensional observations. In this case, all sensors that measure a $d$-dimensional observation that lies inside the $d$-dimensional transmission region transmit the same waveform $s(t)$. As a result, the bandwidth requirement does not depend on\footnote{By contrast, under the TBMA scheme \cite{Mergen_Asymptotic_2007, Liu_Type_2007}, each dimension must be quantized and transmitted to obtain the type of dimension at the FC. While efficient fusion can be done when the features are correlated by whitening or Distributed KLT methods \cite{gastpar2006distributed, nurdin2009reduced, amar2010recursive}, in the worst case the bandwidth requirement grows linearly with $d$ when the dimensions are independent.} $d$. Note that finding the optimal $\Gamma_n$ in terms of minimizing the number of transmissions under reliability constraints might not obey a simple structure, and is likely to require non-convex search algorithms. This issue arises in the TBMA scheme as well, where finding the optimal $K$ quantization values for each dimension is even more complex.

Finally, the scheme can be implemented by dumb simple sensors (oblivious to the observation statistics and without local data processing at sensors). SEEMA simplifies both transmitter and receiver, since the FC receiver can be implemented using a simple AM detector while the sensor requires only an AM transmitter. Unlike LBMA that requires complete knowledge of the observation distributions to compute and transmit the random LLR value by each sensor node, under SEEMA, the sensor nodes only need a few instructions from the FC (i.e., only knowing $A_n$, $\Gamma_n$ is required). Furthermore, transmitting the random LLR values, which suffer from a large dynamic range, might cause signal distortion due to nonlinear effects.

\begin{figure}[htbp]
\centering \epsfig{file=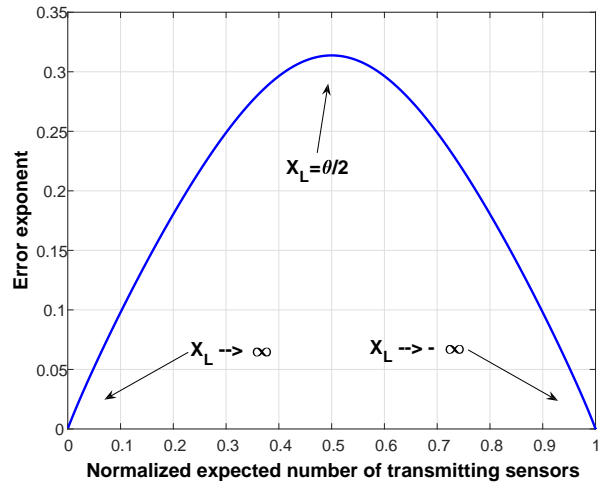,
width=0.5\textwidth}
\caption{Error exponent as a function of the normalized expected number of transmitting sensors. Simulation parameters: $\theta=2$, $\sigma_v^2=1$.}
\label{fig:fig0}
\end{figure}

For instance, when detecting a parameter $\theta$ in AWGN with variance $\sigma_v^2$, the observation distributions are given by $x_n\sim\mathcal{N}(0, \sigma_v^2)$ under $H_0$ and $x_n\sim\mathcal{N}(\theta, \sigma_v^2)$ under $H_1$. A good choice of $\Gamma_n$ is $\Gamma_n=\{x: X_L<x_n<\infty\}$, since the distance between $p_{0,n}$ and $p_{1,n}$ increases in this region. We illustrate this observation in Fig. \ref{fig:fig0} by evaluating the error exponent, defined as the rate function of the error probability as detailed in Section \ref{ssec:LD}, as a function of the normalized expected number of transmitting sensors (which is equal to $P(H_0)p_0+P(H_1)p_1$). The maximal error exponent is achieved at $X_L=\theta/2$, as expected. Setting $X_L>\theta/2$ reduces the expected number of transmissions but concomitantly reduces the error exponent. On the other hand, setting $X_L<\theta/2$ is undesirable because it increases the number of transmissions and decreases the error exponent as well, since that setting $X_L<\theta/2$ decreases the distance between hypotheses due to the single-waveform transmission. For example, if $X_L=-\infty$ all sensors transmit the same waveform and we cannot distinguish between hypotheses.
By contrast, when detecting a normal distributed signal $\theta\sim\mathcal{N}( 0 , \sigma_{\theta}^2 )$ in AWGN, we have $x_n\sim\mathcal{N}( 0 , \sigma_{v}^2 )$ under $H_0$ and $x_n\sim\mathcal{N}( 0 , \sigma_{\theta}^2+\sigma_{v}^2 )$ under $H_1$. Therefore, a good choice of $\Gamma_n$ in this case would be $\Gamma_n=\{x: X_L<|x_n|<\infty\}$.
Determining the transmission region $\Gamma_n$ can be done numerically over $f_{X_n}( x | H_0 ), f_{X_n}( x | H_1 )$
at the FC. Performance can be improved by optimizing the error exponent over the couples $p_{0,n},p_{1,n}$.

\section{Performance Analysis}
\label{sec:performance}

In this section, we analyze the performance of the proposed threshold-based detector (\ref{eq:detector}) in the case of finite $N$ and in the asymptotic regime (where $N\rightarrow\infty$). We first define the notations that will be used in this section. For Bernoulli random variables (r.v) $x,z$ with success probability $q_0$ and $q_1$, respectively, the
Kullback Leibler (KL) divergence between $x,z$ is defined by:
\beq
\label{eq:D_x_y}
\bea{l}
\vspace{0.0cm}
D(x||z)\triangleq D(q_0||q_1)=q_0\log\left(\frac{q_0}{q_1}\right)+(1-q_0)\log\left(\frac{1-q_0}{1-q_1}\right).
\ena
\eeq
Note that under SEEMA, $\textbf{1}_{\Gamma_n}(x_n)$ is a Bernoulli r.v. with success probability $p_{i,n}$
under $H_i$.
Let
\beq
\label{eq:D}
\bea{l} \vspace{0.0cm}
    \overline{D}(p_i||p_j)\triangleq\displaystyle\frac{1}{N}\sum_{n=1}^{N}{D(p_{i,n}||p_{j,n})} \;, \mbox{for\;} i, j=0, 1
\ena
\eeq
denote the average KL divergence across the sensors, and let
\beq
\bea{l}
\label{eq:Lambda}
\displaystyle\Lambda(t)\triangleq\lim_{N\rightarrow\infty}\frac{1}{N}\log E\left\{ e^{Nt\displaystyle y_{_N}} \right\} \;.
\ena
\eeq
Function $\Lambda(t)$ will be used to evaluate the rate function of the detection error by the Gatner-Ellis Theorem, as detailed in Section \ref{ssec:LD}.

The error probability $P_{e,N}$ under SEEMA used in a WSN that contains $N$ sensors is defined by:
\beq
\label{eq:P_e}
\bea{l}
    P_{e,N}=P(H_0)P_N\left(H_0\rightarrow H_1\right)+P(H_1)P_N\left(H_1\rightarrow H_0\right) \;,
\ena
\eeq
where $P_N\left(H_0\rightarrow H_1\right)$ is the probability of declaring $H_1$ when $H_0$ is true (Type-I error probability), and $P_N\left(H_1\rightarrow H_0\right)$ is the probability of declaring $H_0$ when $H_1$ is true (Type-II error probability) in a WSN that contains $N$ sensors. Note that $P_N\left(H_0\rightarrow H_1\right), P_N\left(H_1\rightarrow H_0\right)$ depend on the number of sensors $N$. However, for convenience we often remove the index $N$ and simply write $P\left(H_0\rightarrow H_1\right), P\left(H_1\rightarrow H_0\right)$. We are interested in characterizing the rate at which $P_{e,N}$ approaches zero as $N$ increases.

\subsection{Background on Large Deviations}
\label{ssec:LD}
Throughout this section we use the Large Deviations Principle (LDP) to characterize the limiting behavior of the error probability under SEEMA scheme. Assuming that $P_{e,N}\approx e^{-NI(x)}$, we are interested in evaluating the rate function $I(x)$ (known as the error exponent) of the error probability. To simplify the presentation we assume $Z=1$ in this section. Otherwise, $y_{_N}$ should be replaced by $y_{_N}/Z$. \vspace{0.1cm} \\
\textsl{Definition 1 \cite{Dembo_Large_1998}:}
Let $G^o$, $\bar{G}$ be the interior and closure of a set $G\subset \mathbb{R}$, respectively.
We say that $y_{_1}, ..., y_{_N}$ satisfy the LDP with a rate function $I$ if, for any $G\subset \mathbb{R}$, we
have:
\beq
\label{eq:LDP}
\bea{l}
\vspace{0.1cm}
-\displaystyle\inf_{x\in G^o} I(x) \leq \liminf_{N\rightarrow\infty} \frac{1}{N}\log Pr\left( y_{_N}\in G  \right)
\vspace{0.2cm} \\ \hspace{2cm}
                                   \leq \displaystyle\limsup_{N\rightarrow\infty} \frac{1}{N}\log Pr\left( y_{_N}\in G  \right)
\vspace{0.2cm}
\\ \hspace{5.5cm}
\leq -\displaystyle\inf_{x\in \bar{G}} I(x) \;,
\ena
\eeq
where $I:\mathbb{R}\rightarrow [0,\infty]$.
The effective domain of $I$ is defined by
$D_I\triangleq \left\{x : I(x)<\infty \right\}$. \vspace{0.1cm} \\
In hypothesis testing, $G$ mostly satisfies the I-continuity property \cite{Dembo_Large_1998}:
\begin{center}
$\bea{l}
    \displaystyle\inf_{x\in G^o} I(x)= \displaystyle\inf_{x\in \bar{G}} I(x)\triangleq I_G \;.
\ena$
\end{center}
Then,
\beq
\label{eq:continuity}
\bea{l}
\vspace{0.1cm}
\displaystyle\lim_{N\rightarrow\infty} \frac{1}{N}\log Pr\left( y_{_N}\in G  \right)=-I_G \;.
\ena
\eeq
The Gartner-Ellis Theorem \cite{Dembo_Large_1998} is used throughout this paper to characterize
the rate function:  \\
Let
\beq
\bea{l}
\Lambda_N(t)\triangleq \log E\left\{ e^{ty_{_N}} \right\} \;,
\ena
\eeq
and let
\beq
\bea{l}
\label{eq:Lambda2}
\Lambda(t)\triangleq \displaystyle\lim_{N\rightarrow\infty}\frac{1}{N}\Lambda_N(Nt)=
\displaystyle\lim_{N\rightarrow\infty}\frac{1}{N}\log E\left\{ e^{Nt\displaystyle y_{_N}} \right\} \;.
\ena
\eeq
\emph{Theorem (Gartner-Ellis):}
If $\Lambda(t)$ (\ref{eq:Lambda}) exists as an extended real number, smooth and continuous, then $y_{_1}, ..., y_{_N}$ satisfy the LDP with a rate function
\beq
\bea{l}
\label{eq:Lambda_x}
\Lambda^*(x)=\displaystyle\sup_{t\in \mathbb{R}}{\left( xt- \Lambda(t) \right)}\;, x\in \mathbb{R} \;.
\ena
\eeq
$\Lambda^*(x)$ is the Fenchel-Legendre transform of $\Lambda(t)$.

In this paper we focus on a threshold-based detector for binary hypothesis testing (\ref{eq:detector}). Without loss of generality we assume that $E\left\{y_{_N}|H_1\right\}>E\left\{y_{_N}|H_0\right\}$. We are interested in characterizing the error exponent of the detector. Let $G_0\triangleq\left\{y | y<T\right\}$ and $G_1\triangleq\left\{y | y>T\right\}$ denote the decision regions. The detector decides $H_0$ if $y_{_N}\in G_0$ or decides $H_1$ if $y_{_N}\in G_1$.
Under hypothesis $H_0$, an error occurs if $y_{_N}\in G_1$, thus $G=G_1$ in (\ref{eq:LDP})-(\ref{eq:Lambda_x}). Therefore, $P\left(H_0\rightarrow H_1\right)=Pr\left(y_{_N}\in G_1 | H_0 \right)=Pr\left(y_{_N}>T | H_0 \right)$. Under hypothesis $H_1$, an error occurs if $y_{_N}\in G_0$, thus $G=G_0$ in (\ref{eq:LDP})-(\ref{eq:Lambda_x}). Therefore, $P\left(H_1\rightarrow H_0\right)=Pr\left(y_{_N}\in G_0 | H_1 \right)=Pr\left(y_{_N}<T | H_1 \right)$.
Assume that $\Lambda(t)$ (\ref{eq:Lambda}) exists as an extended real number, smooth and continuous. Then, applying the Gartner-Ellis Theorem to characterize the error exponent of the detector yields:
\beq\label{eq:Gartner_Ellis_threshold}
\bea{l}
\vspace{0.1cm}
-\displaystyle\lim_{N\rightarrow\infty} \frac{1}{N}\log P\left( H_0\rightarrow H_1 \right) \vspace{0.1cm}\\
\hspace{1cm}=-\displaystyle\lim_{N\rightarrow\infty} \frac{1}{N}\log Pr\left(y_{_N}>T | H_0 \right)
=\displaystyle\inf_{x>T}{\Lambda_0^*\left(x\right)}
 \;,\vspace{0.2cm}\\
 -\displaystyle\lim_{N\rightarrow\infty} \frac{1}{N}\log P\left( H_1\rightarrow H_0 \right) \vspace{0.1cm}\\
\hspace{1cm}=-\displaystyle\lim_{N\rightarrow\infty} \frac{1}{N}\log Pr\left(y_{_N}<T | H_1 \right)
=\displaystyle\inf_{x< T}{\Lambda_1^*\left(x\right)} \;,
\ena
\eeq
where $\Lambda^*\left(x\right)=\Lambda_i^*\left(x\right)$ under hypothesis $H_i$ in (\ref{eq:LDP})-(\ref{eq:Lambda_x}).\\
Typically, in hypothesis testing, $T\in \left( E\left\{y_{_N}|H_0\right\} , E\left\{y_{_N}|H_1\right\} \right)$. In this case we have:
\beq\label{eq:Gartner_Ellis_threshold_HT}
\bea{l}
-\displaystyle\lim_{N\rightarrow\infty} \frac{1}{N}\log P\left( H_0\rightarrow H_1 \right) =\Lambda_0^*\left(T\right)
 \;,\vspace{0.2cm}\\
 -\displaystyle\lim_{N\rightarrow\infty} \frac{1}{N}\log P\left( H_1\rightarrow H_0 \right)=\Lambda_1^*\left(T\right) \;.
\ena
\eeq

\subsection{The Case of Equal Channel Gains}
\label{ssec:equal}

We start by analyzing the performance under the no-fading case. To simplify the presentation, we present the results for normalized channels, i.e., $h_n=h_m=1 \;\forall n,m\in\left\{1, ..., N\right\}$. In this case we set $Z=1$ in (\ref{eq:detector}). For $h_n=h_m\neq 1 \;\forall n,m\in\left\{1, ..., N\right\}$ we set $Z=h_n$ in (\ref{eq:detector}), and the analytic results follow by replacing $\sigma^2$ by $\sigma^2/Z^2$.

\textsl{\theorem\label{th:no_fading}{
Assume that the proposed threshold-based detector (\ref{eq:detector}) is implemented. Let $\overline{\Delta}_{0,N}=\overline{D}(p_0||p_1)+\log(\eta)/N$ and $\overline{\Delta}_{1,N}=\overline{D}(p_1||p_0)-\log(\eta)/N$. Let $N_0$ be the minimal number of sensors such that $\overline{\Delta}_{0,N}>0$ and $\overline{\Delta}_{1,N}>0$.
Then:\\
$a\left.\right)$ Consider the case of independent observations under $H_i$. Then, for all $N>N_0$, the error probability is upper bounded by:
\beq
\label{eq:no_fading_inid_Hoeffding}
\bea{l}\vspace{0.0cm}
    P(H_0\rightarrow H_1) \leq \exp\left\{-N\frac{ 2\overline{\Delta}_{0,N}^2}{ \frac{1}{N}\sum_{n=1}^{N}{A_n^2} +4\sigma^2/NE_N}\right\}, \vspace{0.3cm}\\ \vspace{0.0cm}
    P(H_1\rightarrow H_0) \leq \exp\left\{-N\frac{ 2\overline{\Delta}_{1,N}^2}{ \frac{1}{N}\sum_{n=1}^{N}{A_n^2} +4\sigma^2/NE_N}\right\}.
\ena
\eeq
$b\left.\right)$ \emph{(tighter bound (coincides with (\ref{eq:iid_no_fading_err_exponent})) under the conditionally i.i.d. case:)} Consider the case of i.i.d. observations under $H_i$. Let $A\triangleq A_n, \; p_0\triangleq p_{0,n}, \; p_1\triangleq p_{1,n}, \; \Gamma\triangleq \Gamma_n\;, \forall 1\leq n\leq N$ be equal for all sensors. Then, for all $N>N_0$, the error probability is upper bounded by:
\beq
\label{eq:no_fading_iid_Chernoff}
\bea{l}\vspace{0.0cm}
    P(H_0\rightarrow H_1) \\\hspace{1cm}\vspace{0.0cm}
                    \leq \exp\left\{-N\left[ D\left(p_0+\overline{\Delta}_{0,N}/A||p_0\right)-\epsilon_0(N) \right]\right\}, \vspace{0.3cm}\\ \vspace{0.0cm}
    P(H_1\rightarrow H_0) \\\hspace{1cm}
                    \leq \exp\left\{-N\left[ D\left(p_1-\overline{\Delta}_{1,N}/A||p_1\right)-\epsilon_1(N) \right]\right\},
\ena
\eeq
where
\begin{center}
$\bea{l}
\vspace{0.0cm}
\epsilon_0(N)=\frac{\sigma^2}{2NA^2E_N}\log^2\left(1+\frac{\overline{\Delta}_{0,N}/A}{p_0(1-p_0-\overline{\Delta}_{0,N}/A)}\right)
\;, \vspace{0.3cm}\\ \vspace{0.0cm}
\epsilon_1(N)=\frac{\sigma^2}{2NA^2E_N}\log^2\left(1+\frac{\overline{\Delta}_{1,N}/A}{\left(p_1-\overline{\Delta}_{1,N}/A\right)(1-p_1)}\right)
\;.
\ena$
\end{center}
Furthermore, if\footnote{The notation $f(N)=\Omega\left(g(N)\right)$ is used for big Omega notation, i.e., there exist constants $C, N_0>0$ such that for all $N>N_0$ we have $f(N)>Cg(N)$.} $E_N=\Omega\left(N^{\epsilon -1}\right)$, for any $\epsilon>0$, then in the asymptotic regime ($N\rightarrow\infty$) the following holds:\\
$c\left.\right)$ Assume that sensors observations are independent but non-necessarily identically distributed (i.ni.d) under $H_i$. Then, the error exponent under the SEEMA scheme is maximized, and achieves the best error exponent which is obtained by the maximum likelihood detector with respect to the transmitted signal $\textbf{1}_{\Gamma_n}(x_n)$. \\
$d\left.\right)$ Consider the case of non-i.i.d. observations under $H_i$. Assume that $\Lambda(t)$ (\ref{eq:Lambda}) exists as an extended real number, smooth and continuous. Then, $y_{{_N}}$ satisfies the LDP with a rate function: $I_i(x)=\sup_{t\in \mathbb{R}}{\left( xt- \Lambda(t) \right)}\;, x\in \mathbb{R}$, under $H_i$ and the error probability decays exponentially with $N$. Furthermore, if the observations are i.i.d. under $H_i$, then the rate function is given by $I_i(x)=D(x||p_i)$ under $H_i$. The asymptotic error exponent is given explicitly by:
\beq
\label{eq:iid_no_fading_err_exponent}
\bea{l}
\vspace{0.0cm}
-\displaystyle\lim_{N\rightarrow\infty}\frac{1}{N}\log\left(P(H_0\rightarrow H_1)\right)\vspace{0.3cm}\\\hspace{0.2cm}
\vspace{0.0cm}
=-\displaystyle\lim_{N\rightarrow\infty}\frac{1}{N}\log\left(P(H_1\rightarrow H_0)\right) \vspace{0.3cm}\\ \hspace{0.2cm}
=D(p_0+D(p_0||p_1)/A||p_0)
=D(p_1-D(p_1||p_0)/A||p_1)
\;.
\ena
\eeq
}}\\
\emph{Proof:}
We start by proving Statement (a).
Let
\begin{center}
$\tau\triangleq \displaystyle\frac{\log(\eta)}{N}+\frac{1}{N}\sum_{n=1}^{N}{\log\left(\frac{1-p_{0,n}}{1-p_{1,n}}\right)}$.
\end{center}
Note that after algebraic manipulations we have:
\begin{center}
$\bea{l}
\displaystyle\tau=\frac{1}{N}\sum_{n=1}^{N}{p_{0,n}A_n}+\overline{\Delta}_{0,N} \;,
\ena$
\end{center}
where $\frac{1}{N}\sum_{n=1}^{N}{p_{0,n}A_n}$ is the expectation of $y_{_N}$ under $H_0$.
Applying the Chernoff bound yields:
\begin{center}
$\bea{l}
Pr\left(
\displaystyle \frac{1}{N}\sum_{n=1}^{N}{A_n\textbf{1}_{\Gamma_n}(x_n)}+\tilde{w}>\tau
\right) \vspace{0.3cm} \\ \hspace{0.2cm}
\displaystyle\leq e^{-sN\tau}\prod_n E\left[\exp\left\{A_n\textbf{1}_{\Gamma_n}(x_n)\right\}\right]E\left[\exp\left\{w/\sqrt{E_N}\right\}\right].
\ena$
\end{center}
Since $A_n$ is bounded by the construction of the transmission scheme, using algebraic manipulations as in the construction of the Hoeffding bound yields:
\begin{center}
$\bea{l}
Pr\left(
\displaystyle \frac{1}{N}\sum_{n=1}^{N}{A_n\textbf{1}_{\Gamma_n}(x_n)+\tilde{w}}>\tau
\right) \vspace{0.3cm} \\ \hspace{2cm}
\leq\exp\left\{-N\frac{ 2\overline{\Delta}_{0,N}^2}{\frac{1}{N}\sum_{n=1}^{N}{A_n^2+4\sigma^2/NE_N}}\right\},
\ena$
\end{center}
where we used the fact that $\tilde{w}\sim subG(\sigma^2/NE_N)$, so that $E\left[e^{s\tilde{w}}\right]\leq e^{s^2\sigma^2/2E_N}$. The second term in (\ref{eq:no_fading_inid_Hoeffding}) can be developed similarly under $H_1$. \vspace{0.2cm}\\
Next, we prove Statement (b). Rewriting (\ref{eq:detector}) with $Z=1$ yields:
\beq
\label{eq:2.b1}
\bea{l}
\displaystyle\sum_{n=1}^{N}{\textbf{1}_{\Gamma}(x_n)}+w'>N\frac{\displaystyle\log(\eta)/N+\log\left(\frac{1-p_0}{1-p_1}\right)}{A}
 \;,
\ena
\eeq
where $w'\sim subG(\sigma^2/(E_N A^2))$. \vspace{0.3cm}\\

Let
\begin{center}
$\displaystyle\tau'\triangleq\tau/A=\left(\log(\eta)/N+\log\left(\frac{1-p_0}{1-p_1}\right)\right)/A$.
\end{center}
be a threshold normalized by $A=\log\frac{(1-p_0)p_1}{(1-p_1)p_0}$ used in the detector (\ref{eq:detector}).
Since the observations are conditionally i.i.d., after algebraic manipulations we have:
\beq
\label{eq:tau}
\bea{l}
\displaystyle\tau'=p_0+\frac{\overline{\Delta}_{0,N}}{A}.
\ena
\eeq
Next, applying the Chernoff bound and using the i.i.d. property yields for all $t\geq 0$:
\beq
\label{eq:2.b2}
\bea{l}
Pr\left(
\displaystyle\sum_{n=1}^{N}{\textbf{1}_{\Gamma}(x_n)}+w'>N\tau'
\right) \vspace{0.3cm}\\ \hspace{0.2cm}
\leq \displaystyle\left[ \frac{E\left\{e^{t\cdot\textbf{1}_{\Gamma}(x_n)}\right\}}{e^{t\tau}}  \right]^N E\left\{e^{tw'}\right\}  \vspace{0.3cm}\\ \hspace{0.2cm}
\leq \displaystyle\left[ \left(e^t\right)^{-\left(p_0+\frac{\overline{\Delta}_{0,N}}{A}\right)}\left(p_0e^t+(1-p_0)\right)\right]^N
e^{t^2 \frac{\sigma^2}{2 E_N A^2}} \;.
\ena
\eeq
Since the bound holds for all $t>0$, by placing $t=\log\frac{(1-p_0)\left(p_0+\frac{\overline{\Delta}_{0,N}}{A}\right)}{(1-p_0-\frac{\overline{\Delta}_{0,N}}{A})p_0}>0$
(or $e^t=\frac{(1-p_0)\left(p_0+\frac{\overline{\Delta}_{0,N}}{A}\right)}{(1-p_0-\frac{\overline{\Delta}_{0,N}}{A})p_0}$) in (\ref{eq:2.b2}), we obtain:
\beq
\label{eq:chernoff_placing_t}
\bea{l}
Pr\left(
\displaystyle\sum_{n=1}^{N}{\textbf{1}_{\Gamma}(x_n)}+w'>N\tau'
\right) \vspace{0.3cm}\\ \hspace{0.2cm}
\leq\displaystyle\left[ \left(\frac{(1-p_0)\left(p_0+\frac{\overline{\Delta}_{0,N}}{A}\right)}{(1-p_0-\frac{\overline{\Delta}_{0,N}}{A})p_0}\right)^{-\left(p_0+\frac{\overline{\Delta}_{0,N}}{A}\right)}
\times \right.\vspace{0.2cm}\\\left.\hspace{0.4cm}
\displaystyle\left(\frac{(1-p_0)\left(p_0+\frac{\overline{\Delta}_{0,N}}{A}\right)}{(1-p_0-\frac{\overline{\Delta}_{0,N}}{A})}+(1-p_0)\right)\right]^N
\times\vspace{0.3cm}\\\hspace{0.4cm}
\displaystyle\exp\left\{\displaystyle\frac{\sigma^2}{2 E_N A^2}\log^2\frac{(1-p_0)\left(p_0+\frac{\overline{\Delta}_{0,N}}{A}\right)}{(1-p_0-\frac{\overline{\Delta}_{0,N}}{A})p_0}\right\} \;.
\ena
\eeq
We can rewrite the RHS of (\ref{eq:chernoff_placing_t}) as:
\beq
\label{eq:chernoff_placing_t_RHS}
\bea{l}
\displaystyle\left[
\left(\frac{p_0+\frac{\overline{\Delta}_{0,N}}{A}}{p_0}\right)^{-\left(p_0+\frac{\overline{\Delta}_{0,N}}{A}\right)}
\times \right.\vspace{0.2cm}\\\left.\hspace{0.4cm}
\displaystyle\left(\frac{1-p_0}{1-p_0-\frac{\overline{\Delta}_{0,N}}{A}}\right)^{-\left(p_0+\frac{\overline{\Delta}_{0,N}}{A}\right)}
\displaystyle\left(\frac{1-p_0}{1-p_0-\frac{\overline{\Delta}_{0,N}}{A}}\right)\right]^N
\times\vspace{0.3cm}\\\hspace{0.4cm}
\displaystyle\exp\left\{\frac{\sigma^2}{2 E_N A^2}\log^2\left(1+\frac{\overline{\Delta}_{0,N}/A}{\left(1-p_0-\frac{\overline{\Delta}_{0,N}}{A}\right)p_0}\right)\right\}
\vspace{0.2cm}\\\hspace{0.4cm}%
=\displaystyle\left[
\left(\frac{p_0+\frac{\overline{\Delta}_{0,N}}{A}}{p_0}\right)^{-\left(p_0+\frac{\overline{\Delta}_{0,N}}{A}\right)}
\times \right.\vspace{0.2cm}\\\left.\hspace{2cm}
\displaystyle\left(\frac{1-p_0}{1-p_0-\frac{\overline{\Delta}_{0,N}}{A}}\right)^{1-\left(p_0+\frac{\overline{\Delta}_{0,N}}{A}\right)}
\times\vspace{0.3cm}\right.\vspace{0.2cm}\\\hspace{0.5cm}\left.
\displaystyle\exp\left\{\frac{\sigma^2}{2 N E_N A^2}\log^2\left(1+\frac{\overline{\Delta}_{0,N}/A}{\left(1-p_0-\frac{\overline{\Delta}_{0,N}}{A}\right)p_0}\right)\right\} \right]^N
\;.
\ena
\eeq
Finally, taking logarithm and exponent yields (\ref{eq:no_fading_iid_Chernoff}).  \vspace{0.2cm}\\
Next, we prove Statement (c). Let
\begin{center}
$L_n=\log\left(p(\textbf{1}_{\Gamma_n}(x_n)|H_1)/p(\textbf{1}_{\Gamma_n}(x_n)|H_0)\right)$
\end{center}
be the log-likelihood ratio of sensor $n$ regarding the r.v. $\textbf{1}_{\Gamma_n}(x_n)$.
The observation $\textbf{1}_{\Gamma_n}(x_n)$ that is used in the SEEMA scheme has pmf:
\begin{center}
$p\left(\textbf{1}_{\Gamma_n}(x_n)|H_j\right)
=p_{j,n}^{\textbf{1}_{\Gamma_n}(x_n)}(1-p_{j,n})^{1-\textbf{1}_{\Gamma_n}(x_n)}$.
\end{center}
Then, we get,
\begin{center}
$\bea{l}
L_n=\textbf{1}_{\Gamma_n}(x_n)\log\left(\frac{p_{1,n}}{p_{0,n}}\right)
    +(1-\textbf{1}_{\Gamma_n}(x_n))\log\left(\frac{1-p_{1,n}}{1-p_{0,n}}\right) .
\ena$
\end{center}
An optimal ML detector decides $H_1$ if
\begin{center}
$\displaystyle\sum_{n=1}^{N}{L_n}>\log(\eta)$.
\end{center}
Therefore,
\beq
\label{App:eq_3_1}
\bea{l}
\displaystyle\sum_{n=1}^{N}{L_n}=\sum_{n=1}^{N}{\textbf{1}_{\Gamma_n}(x_n) A_n+\log\left(\frac{1-p_{1,n}}{1-p_{0,n}}\right)}>\log(\eta)
 \;.
\ena
\eeq
Otherwise, it decides $H_0$. \\
Rearranging (\ref{App:eq_3_1}) yields (\ref{eq:detector}) in a no-fading and noise-free channel scenario.
Then, (\ref{eq:detector}) can be rewritten as:
\begin{center}
$\bea{l}
    \displaystyle\sum_{n=1}^{N}{L_n}+w'
    >\log(\eta)
 \;,
\ena$
\end{center}
where $w'\sim subG(\sigma^2/E_N)$.
Since $E\left[e^{tw'}\right]\leq e^{\frac{t^2\sigma^2}{2E_N}}$, we have:
\begin{center}
$\bea{l}
\displaystyle\frac{1}{N}\log E\left\{e^{t\left(\sum_{n=1}^{N}{L_n}+w'\right)}\right\} \vspace{0.3cm}\\
=\displaystyle\frac{1}{N}\log \prod_{n=1}^{N}{E\left\{e^{t L_n}\right\}}
                         +O(1/(E_N N))   \vspace{0.3cm}\\
\longrightarrow \displaystyle
\frac{1}{N}\sum_{n=1}^{N}
{\log E\left(\frac{p(\textbf{1}_{\Gamma_n}(x_n)|H_1)}{p(\textbf{1}_{\Gamma_n}(x_n)|H_0)}\right)^t}
= \Lambda(t) \vspace{0.3cm} \\
\hspace{3cm}
\mbox{as $N\rightarrow\infty$ and $E_N=\Omega(N^{\epsilon-1})$}
\ena$
\end{center}
Since we obtained the rate function of the optimal noise-free LLR test that minimizes the error probability, the statement follows. \vspace{0.2cm}\\
Next, we prove Statement (d). The Gartner-Ellis conditions are assumed to be satisfied.
Note that after algebraic manipulations of the threshold $\tau$ we have:
\begin{center}
$\bea{l}
\displaystyle\tau=\frac{1}{N}\sum_{n=1}^{N}{p_{0,n}A_n}+\overline{\Delta}_{0,N}
                 =\frac{1}{N}\sum_{n=1}^{N}{p_{1,n}A_n}-\overline{\Delta}_{1,N}\;,
\ena$
\end{center}
where $\frac{1}{N}\sum_{n=1}^{N}{p_{0,n}A_n}$ and $\frac{1}{N}\sum_{n=1}^{N}{p_{1,n}A_n}$ are the expectations of $y_{_N}$ under $H_0$ and $H_1$, respectively. Since an error under $H_0$ occurs when $y_{_N}>\tau=E[y_{_N}|H_0]+\overline{\Delta}_{0,N}$, an error under $H_1$ occurs when $y_{_N}<\tau=E[y_{_N}|H_1]-\overline{\Delta}_{1,N}$, and $\overline{\Delta}_{0,N}\;,\;\overline{\Delta}_{1,N}$ are strictly positive, the error probability decays exponentially with $N$ since the Gartner-Ellis conditions are satisfied by the assumption.

We continue by proving the statement under the i.i.d. observations case. Rewriting (\ref{eq:detector})  yields:
\beq
\label{eq:2.c1}
\bea{l}
   y'\triangleq\displaystyle\frac{1}{N}\sum_{n=1}^{N}{\textbf{1}_{\Gamma_n}(x_n)}+w'
   >\frac{\displaystyle\log(\eta)/N+\log\left(\frac{1-p_0}{1-p_1}\right)}{A}
 \;,
\ena
\eeq
where $w'\sim subG(\sigma^2/(N^2E_N A^2))$. \\
We need to show that $y'$ satisfies the LDP.\\
Let
\begin{center}
$\Lambda_N(t)\triangleq \log E\left\{ e^{ty'} \right\} \;,$
\end{center}
and let
\begin{center}
$\Lambda(t)\triangleq \displaystyle\lim_{N\rightarrow\infty}\frac{1}{N}\Lambda_N(Nt)=
\displaystyle\lim_{N\rightarrow\infty}\frac{1}{N}\log E\left\{ e^{Nty'} \right\}$.
\end{center}
The Gartner-Ellis Theorem states that if $\Lambda(t)$ exists as an extended real number, smooth and continuous, then $y'$ satisfies the LDP with a rate function
\begin{center}
$\Lambda^*(x)=\displaystyle\sup_{t\in \mathbb{R}}{\left( xt- \Lambda(t) \right)}\;, x\in \mathbb{R}$ \;,
\end{center}
dubbed the Fenchel-Legendre transform of $\Lambda(t)$.
Due to the i.i.d. property, and the fact that $\frac{1}{N}\log E\left\{ e^{Ntw'} \right\}\leq\frac{\sigma^2t^2}{2NE_N}$, we have:
\begin{center}
$\bea{l}
\frac{1}{N}\log E\left\{ e^{Nty'} \right\} \vspace{0.3cm}\\
=\frac{1}{N}\log \left[E\left\{ e^{t\cdot\textbf{1}_{\Gamma_n}(x_n)} \right\}\right]^N
+\frac{1}{N}\log E\left\{ e^{Ntw'} \right\} \vspace{0.3cm}\\
\longrightarrow \log E\left\{ e^{t\cdot\textbf{1}_{\Gamma_n}(x_n)} \right\} \;,
\mbox{as $N\rightarrow\infty$ and $E_N=\Omega(N^{\epsilon-1})$}  \vspace{0.3cm}\\
=\log \left( p_0 e^t + 1-p_0  \right)=\Lambda(t) \;.
\ena$
\end{center}
$\Lambda(t)$ is smooth and continuous. Hence, $y'$ satisfies the LDP with a rate function $\Lambda^*(x)=\sup_{t\in \mathbb{R}}{\left( xt- \Lambda(t) \right)}$. Differentiating and equating the derivative to zero, yields:
\begin{center}
$\Lambda^*(x)=x\log\frac{x}{p_0}+(1-x)\log\frac{1-x}{1-p_0}=D(x||p_0) \;.$
\end{center}
Finally, similar to (\ref{eq:tau}), the RHS of (\ref{eq:2.c1}) satisfies $\left[\displaystyle\log(\eta)/N+\log\left((1-p_0)/(1-p_1)\right)\right]/A\rightarrow p_0+D(p_0||p_1)/A$ as $N\rightarrow\infty$. Hence, the theorem follows.
\vspace{0.0cm}
\hfill $\square$ \vspace{0.3cm}

\subsection{The Case of Fading Channels}
\label{ssec:fading}

We next deal with the case where $h_n=h_m$ may not hold. Note that in the SEEMA transmission scheme, sensors transmit the analog waveform without directly correcting the channel gain (although correcting the phase is assumed to avoid zero-mean channels). As a result, the received signals at the FC are multiplied by random channel gains. Nevertheless, this transmission scheme is applicable to many common applications: (i) The case where the channel gain is not corrected at the transmitter to make the scheme robust against changes in the channel statistics. Thus, the average transmission energy of the signal is determined according to the observations statistics purely to satisfy the average energy constraints. This transmission scheme is very simple to implement and is generally preferred in WSNs with a mobile access point \cite{Tong_Sensor_2003}; for instance, where the channel statistics can vary rapidly and are not available at the sensors. Note that correcting the channel phase (by transmitting a signal with the complex conjugate channel phase) is assumed to avoid zero-mean channels, as was done in \cite{Mergen_Type_2006, Mergen_Asymptotic_2007, cohen2013performance, wimalajeewa2015wireless}; however, the signal energy is not affected by this operation. Correcting the channel phase can be done by transmitting a pilot signal by the FC before the sensor transmissions to estimate the channel phase \cite{Chen_Transmission_2007, Cohen_TOP_2010}. In fact, estimating the channel phase by sensors with an estimation error of less than $\pi/4$ is sufficient to correct the phase at the transmitter to guarantee positive $I, Q$ components at the receiver. (ii) The case where sensors exploit the channel state to correct the fading effect (for instance, by dividing the signal amplitude at the transmitters by the channel state to obtain identical channels at the FC). However, due to channel estimation errors, the signal amplitudes are still multiplied by random gains at the FC. (iii) The case where the sensors adapt their transmission power according to the channel state to obtain discrete channels at the FC. For example, consider a transmission scheme where each sensor transmits its waveform divided by the channel state only if the channel gain is greater than a predetermined threshold (to satisfy a power constraint). Otherwise, the sensor does not transmit (to save energy). As a result, the FC receives the transmitted signals multiplied by $1$ (good channel) with probability $p$, where $p$ is the probability that the channel gain is greater than the predetermined threshold. All other signals are multiplied by $0$ (bad channel) with probability $1-p$. This scenario is known as a transmission scheme over ON/OFF fading channels.

As discussed in Section \ref{ssec:LD}, we need to set $Z$ in (\ref{eq:detector}) such that $\tau\in \left( E\left\{\frac{y_{_N}}{Z}|H_0\right\} , E\left\{\frac{y_{_N}}{Z}|H_1\right\} \right)$ to achieve the desired decay of error, where $\tau=\frac{\log(\eta)}{N}+\frac{1}{N}\sum_{n=1}^{N}{\log\left(\frac{1-p_{0,n}}{1-p_{1,n}}\right)}$ is the detector's threshold in (\ref{eq:detector}). When the channel gains are i.i.d. across sensors, we have $\mu_{h,n}=\mu_{h,m}$ for all $n,m\in\left\{1, ..., N\right\}$. In this case, by setting\footnote{Note that letting the sensors transmit a few pilot signals which are coherently aggregated at the FC yields a good estimate of $\mu_{h,n}$ under the i.i.d. fading channels case.} $Z=\mu_{h,n}$, and using algebraic manipulations we have:
\begin{center}
$\bea{l}
\displaystyle\tau=\frac{1}{N}\sum_{n=1}^{N}{p_{0,n}A_n}+\overline{\Delta}_{0,N}=\frac{1}{N}\sum_{n=1}^{N}{p_{1,n}A_n}-\overline{\Delta}_{1,N} \;,
\ena$
\end{center}
where $\frac{1}{N}\sum_{n=1}^{N}{p_{0,n}A_n}$, $\frac{1}{N}\sum_{n=1}^{N}{p_{1,n}A_n}$, are the expectations of $y_{{_N}}/Z$ under hypotheses $H_0, H_1$, respectively, and $\overline{\Delta}_{0,N}=\overline{D}(p_0||p_1)+\log(\eta)/N$ and $\overline{\Delta}_{1,N}=\overline{D}(p_1||p_0)-\log(\eta)/N$. As a result, there exists a number $N_0$ of sensors such that for all $N>N_0$ we have: $\tau\in \left( E\left\{\frac{y_{_N}}{Z}|H_0\right\} , E\left\{\frac{y_{_N}}{Z}|H_1\right\} \right)$. When the channel gains are not i.i.d. across sensors we assume that $Z$ satisfies the desired property
\footnote{Note that satisfying $\tau\in \left( E\left\{y_{_N}/Z|H_0\right\} , E\left\{y_{_N}/Z|H_1\right\} \right)$ does not require high accuracy when the FC estimates the expectations $E\left\{y_{_N}/Z|H_0\right\}, E\left\{y_{_N}/Z|H_1\right\}$. We point out that such (and even more complex) learning mechanisms are required in other schemes as well, such as computing a complex rate function that depends on the channel distribution of each sensor node for TBMA \cite{Mergen_Asymptotic_2007}.
}. \vspace{0.2cm}

\noindent \emph{Assumption A1:} Let $\Delta_{0,N}=\tau-E\left\{\frac{y_{_N}}{Z}|H_0\right\}$, and $\Delta_{1,N}=E\left\{\frac{y_{_N}}{Z}|H_1\right\}-\tau$. There exists a number $N_0$ of sensors such that for all $N>N_0$ we have: $\Delta_{0,N}>0$ and $\Delta_{1,N}>0$.\vspace{0.2cm}

In contrast to the case of identical channels and i.i.d. observations, SEEMA does not achieve the centralized error exponent when transmitting using fading channels. However, exponential decay is still obtained when transmitting with energy $E_N\sim N^{-1}$ as shown below.

\textsl{\theorem\label{th:fading_error_exponent}{
Assume that the proposed threshold-based detector (\ref{eq:detector}) is implemented and Assumption A1 holds. Then:\\
$a\left.\right)$ Consider the case of independent observations under $H_i$ and independent channel gains which are upper bounded by $|h_n/Z|<h_{max}$ for all $n\in\left\{1, ..., N\right\}$. Let $N_0$ be the minimal number of sensors such that $\Delta_{0,N}>0$ and $\Delta_{1,N}>0$. Then, for all $N>N_0$, the error probability is upper bounded by:
\beq
\label{eq:iid_fading_Hoeffding}
\bea{l}\vspace{0.1cm}
    P(H_0\rightarrow H_1) \leq \exp\left\{-N\frac{ 2\Delta_{0,N}^2}{ h_{max}^2\frac{1}{N}\sum_{n=1}^{N}{A_n^2} +4\sigma^2/(NE_N Z^2)}\right\}, \\ \vspace{0.1cm}
    P(H_1\rightarrow H_0) \leq \exp\left\{-N\frac{ 2\Delta_{1,N}^2}{ h_{max}^2\frac{1}{N}\sum_{n=1}^{N}{A_n^2} +4\sigma^2/(NE_N Z^2)}\right\}.
\ena
\eeq
Furthermore, if $E_N=\Omega\left(N^{\epsilon -1}\right)$, for any $\epsilon>0$, then in the asymptotic regime ($N\rightarrow\infty$) the following holds:\\
$b\left.\right)$ Consider the case of non-i.i.d. observations and non-i.i.d. fading channels under $H_i$. Assume that $\Lambda(t)$ (\ref{eq:Lambda}) exists as an extended real number, smooth and continuous. Then, $y_{{_N}}$ satisfies the LDP with a rate function: $I_i(x)=\sup_{t\in \mathbb{R}}{\left( xt- \Lambda(t) \right)}\;, x\in \mathbb{R}$, under $H_i$ and the error probability decays exponentially with $N$.
Furthermore, consider the case of i.i.d. observations under $H_i$, i.i.d. channel gains, and assume that the moment generating function of the channels is finite $E\left\{e^{th_n}\right\}<\infty$. Let $A\triangleq A_n, \; p_0\triangleq p_{0,n}, \; p_1\triangleq p_{1,n}, \; \Gamma\triangleq \Gamma_n\;, \forall 1\leq n\leq N$ be equal for all sensors. Then, $y_{{_N}}$ satisfies the LDP with a rate function:  $I_i(x)=\sup_{t\in \mathbb{R}}{\left( xt-\Lambda(t)\right)}$, where $\Lambda(t)=\log\left(p_{i} E\left\{e^{ \frac{tA h_n}{Z}}\right\}+1-p_i\right)$ under $H_i$, $i=0, 1$, and the error probability decays exponentially with $N$.
}}\vspace{0.2cm}\\
\emph{Proof:}
We start by proving Statement (a) under hypothesis $H_0$.
Since an error occurs under $H_0$ when $y_{{_N}}>\tau$, we can apply the Chernoff bound to upper bound the error probability:
\begin{center}
$\bea{l}
Pr\left(
\displaystyle \frac{1}{N}\sum_{n=1}^{N}{h_nA_n\textbf{1}_{\Gamma_n}(x_n)/Z}+\tilde{w}/Z>\tau
\right) \vspace{0.3cm} \\ \hspace{0.5cm}
\displaystyle\leq e^{-sN\tau}\prod_n E\left[\exp\left\{\frac{h_nA_n\textbf{1}_{\Gamma_n}(x_n)}{Z}\right\}\right]\times \vspace{0.3cm} \\ \hspace{3cm} E\left[\exp\left\{\frac{w}{Z\sqrt{E_N}}\right\}\right].
\ena$
\end{center}
Since $Pr(h_n A_n\textbf{1}_{\Gamma_n}(x_n)/Z\in [-h_{max} A_n,h_{max} A_n])=1$ and Assumption A1 holds, using algebraic manipulations as in the construction of the Hoeffding bound yields:
\begin{center}
$\bea{l}
Pr\left(
\displaystyle \frac{1}{N}\sum_{n=1}^{N}{h_nA_n\textbf{1}_{\Gamma_n}(x_n)/Z+\tilde{w}/Z}>\tau
\right) \vspace{0.3cm} \\ \hspace{2cm}
\leq\exp\left\{-N\frac{ 2\Delta_{0,N}^2}{h_{max}^2\frac{1}{N}\sum_{n=1}^{N}{A_n^2+4\sigma^2/(NE_NZ^2)}}\right\},
\ena$
\end{center}
where we used the fact that $\tilde{w}/Z\sim subG(\sigma^2/E_NZ^2)$, so that $E\left[e^{s\tilde{w}/Z}\right]\leq e^{s^2\sigma^2/2E_NZ^2}$. The second term in (\ref{eq:iid_fading_Hoeffding}) can be developed similarly under $H_1$. \vspace{0.2cm}\\
Statement (b) follows by similar steps as in the proof of Theorem \ref{th:no_fading}.d under the non-i.i.d. case by taking expectation with respect to the channel gain. Under the i.i.d. case, the statement follows by using the fact that we have: $\Lambda(t)=\log\left(p_i E\left\{e^{ \frac{tA h_n}{Z}}\right\}+1-p_i\right)$ due to the i.i.d. property.
\vspace{0.0cm}
\hfill $\square$ \vspace{0.3cm}

\begin{remark}
Note that the noise decay in (\ref{eq:y}) implies that the detector's performance could be improved by increasing the number of sensors in the network without increasing the total transmission energy. Theorems \ref{th:no_fading}, \ref{th:fading_error_exponent} characterize the decay rate of the error probability, which holds if the transmitted energy satisfies $E_N=\Omega\left(N^{\epsilon -1}\right)$. These results provides important design principles for detection under resource constraints.
\end{remark}

\subsection{Explicit Analysis in Common Non-I.I.D. Scenarios}

For non-i.i.d. observations, the conditions on $\Lambda(t)$ are used to apply the Gartner-Ellis Theorem to obtain the rate function. Next, we illustrate common cases in WSNs when the conditions hold. \vspace{0.2cm}

\subsubsection{The Case of Local Conditionally I.I.D. Observations}

First, consider a common scenario where sensors are located in $K$ different areas, where a set of $\mathcal{N}_k$ sensors with cardinality $N(k)$ is located in area $k$, and their observations are independent but not necessarily identically distributed under $H_i$. However, sensor observations in the same area $k$ $(1\leq k\leq K)$ are assumed to be i.i.d. under $H_i$ (due to the small geographical distance between them). Let $\tilde{A}_k=A_n$ for all sensors (say $n$) in area $k$. Assume that the sensor deployment process follows a multinomial distribution with probabilities $p_1, ..., p_K$, where $\sum_{i=1}^K p_i=1$. Specifically, when deploying the sensors in the field, each sensor has a probability $p_i$ of being located in area $i$. Assume that the channel gains $h_n$ are i.i.d. within each region, and $E_N=\Omega\left(N^{\epsilon-1}\right)$, for fixed $\epsilon>0$. In this scenario, we have:
\beq
\label{eq:lambda_local_iid}
\bea{l}
\vspace{0.0cm} \displaystyle\frac{1}{N}\log\left(E\left\{e^{Nty}\right\}\right) \vspace{0.3cm}\\
\displaystyle=\frac{1}{N}\log\left(E\left\{e^{t\sum\limits_{n=1}^{N}{h_n A_n\textbf{1}_{\Gamma_n}(x_n)}+Nt\tilde{w}}\right\}\right) \vspace{0.3cm}\\
\displaystyle=\frac{1}{N}\log\left(E\left\{\prod_{k=1}^{K} e^{t\sum\limits_{n\in\mathcal{N}_k}{h_n \tilde{A}_k\textbf{1}_{\Gamma_n}(x_n)}}\right\}\right)
\vspace{0.3cm}\\\hspace{4cm}
+\displaystyle\frac{1}{N}\log E\left\{e^{tN\tilde{w}}\right\}
 \vspace{0.3cm}\\
\displaystyle=\sum_{k=1}^K\frac{N(k)}{N}\log\left(E\left\{e^{t{h \tilde{A}_k\textbf{1}_{\Gamma_n}(x)}}\right\}\right)
+O(1/(NE_N))
 \vspace{0.3cm}\\
\displaystyle\longrightarrow E_{\mathcal{N}_k}\left\{\log \left(p_i E_h\left\{ e^{th_n \tilde{A}_k} \right\}+1-p_i\right)\right\}=\Lambda(t) \;, \vspace{0.3cm}\\\hspace{3cm}
\mbox{as $N\rightarrow\infty$, and $E_N=\Omega\left(N^{\epsilon-1}\right)$}  , \vspace{0.2cm}
\ena
\eeq
under $H_i$, $i=0, 1$, by the law of large numbers. The first expectation is with respect to the sensor deployment process, and the second expectation is with respect to the channel gain given that sensors lie in the set $\mathcal{N}_k$.
In this case, we only need to require that the moment generating function of the channel fading be finite to guarantee that Statement (b) in Theorem \ref{th:fading_error_exponent} holds. For example, the detection performance over a Rayleigh fading channel with received power $P_r=E\left\{|h|^2\right\}$ is evaluated by setting:
\begin{center}
$\bea{l}
\displaystyle E_h\left\{ e^{th_n \tilde{A}_k}\right\} \vspace{0.2cm}\\ \hspace{0.2cm} \displaystyle=1+\frac{1}{\sqrt{2}}\frac{\sqrt{P_r}}{\tilde{A}_k}te^{\frac{P_rt^2}{4\tilde{A}_k^2}}\sqrt{\frac{\pi}{2}}\left(erf\left(\sqrt{\frac{P_rt}{4\tilde{A}_k^2}}\right)+1\right)
\ena$
\end{center}
in $\Lambda(t)$, and solving $I_i(x)=\sup_{t\in \mathbb{R}}{\left( xt-\Lambda(t)\right)}$ under $H_i$, for detection threshold $x$.
\vspace{0.2cm}

\subsubsection{The Case of Spatially Correlated Markovian Observations}
\label{ssec:special_Markovian}
In this section we examine the case of correlated Markovian observations across sensors, in which the random variables $\textbf{1}_{\Gamma_n}(x_n)$ form a Markov chain across $n=1, ..., N$ (i.e., a one-dimensional field), i.e., a spatial Gilbert-Elliot model.
Specifically, let $\pi_m(i,j)$ be the transition probability to observe $\textbf{1}_{\Gamma_n}(x_n)=j$ given that the neighbor node observes $\textbf{1}_{\Gamma_n}(x_{n-1})=i$, for $i, j\in\left\{0,1\right\}$ under $H_m$. For convenience, we will neglect subscript $m$ during the analysis.
Let $\Pi=\left\{\pi(i,j)\right\}_{i,j=0}^1$ be the transition probability matrix of $\textbf{1}_{\Gamma_n}(x_n)$ across the sensors. Let $P_{s}^{\pi}$ be the Markov probability measure with the initial state $s\in\left\{0,1\right\}$:
\begin{center}
$\bea{l}
\displaystyle
P_{s}^{\pi}(\textbf{1}_{\Gamma_1}(x_1)=\alpha_{1}, ..., \textbf{1}_{\Gamma_N}(x_N)= \alpha_{N}) \vspace{0.2cm} \\ \hspace{1cm} \displaystyle
=\pi(s,\alpha_{1})\prod_{n=1}^{N-1}{\pi(\alpha_{n}, \alpha_{{n+1}})} \;,
\ena$
\end{center}
where $\alpha_{n}\in\left\{0,1\right\}$ for each sensor $n$.

Let $\tilde{y}\triangleq y/A$. Next, we compute $\lim_{N\rightarrow\infty}{\frac{1}{N} \log E_s^{\pi}\left\{e^{Nt\tilde{y}}\right\}}$, where $E_s^{\pi}\left\{e^{Nt\tilde{y}}\right\}$ is the expectation of $e^{Nt\tilde{y}}$ with respect to $P_{s}^{\pi}(\textbf{1}_{\Gamma_1}(x_1)=\alpha_{k_1}, ..., \textbf{1}_{\Gamma_N}(x_N)= \alpha_{k_N})$, to use the Gartner-Ellis Theorem.
Note that under $H_m$ we have:
\begin{center}
$\bea{l}
\frac{1}{N}\Lambda_N(Nt) \vspace{0.1cm} \vspace{0.2cm} \\
=\frac{1}{N}\log E_s^{\pi}\left\{e^{Nt\tilde{y}}\right\}=\frac{1}{N}\log E_s^{\pi}\left\{e^{t\sum_{n=1}^{N}{h_n \textbf{1}_{\Gamma_n}(x_n)}+Nt\tilde{w}}\right\} \vspace{0.1cm} \vspace{0.2cm} \\
=\frac{1}{N}\displaystyle\log\hspace{-0.2cm} \sum_{\alpha_1=0,1}\hspace{-0.2cm}{\cdots\hspace{-0.2cm} \sum_{\alpha_N=0,1}{\hspace{-0.2cm} P_{s}^{\pi}(\textbf{1}_{\Gamma_1}(x_1)=\alpha_{1}, ..., \textbf{1}_{\Gamma_N}(x_N)= \alpha_{N})} } \times \vspace{0.2cm} \\\hspace{5.5cm}
\displaystyle\prod_{n=1}^{N}{E_{h}\left\{ e^{t h_n \alpha_{n}} \right\}  } \vspace{0.1cm} \vspace{0.2cm} \\ \hspace{6cm}
+\displaystyle O(1/(NE_N)) \vspace{0.2cm}
\ena$
\end{center}
\begin{center}
$\bea{l}
=\frac{1}{N}\displaystyle\log \sum_{\alpha_1=0,1}\cdots \sum_{\alpha_N=0,1}\pi(s,\alpha_{1})E_{h}\left\{ e^{t h_1 \alpha_{1}} \right\}
\times\cdots \vspace{0.2cm} \\ \hspace{3cm} \times
\pi(\alpha_{{N-1}},\alpha_{N})E_{h}\left\{ e^{t h_N \alpha_{N}} \right\}  \vspace{0.2cm} \\ \hspace{6cm}
+\displaystyle O(1/(NE_N)) \vspace{0.2cm} \\
\longrightarrow
\frac{1}{N}\displaystyle\log  \sum_{\alpha_N=0,1} \left( \Pi_{t} \right)^N\left(s,\alpha_{N} \right) \;,
\vspace{0.2cm} \\ \hspace{3cm}
\mbox{as $N\rightarrow\infty$ and $E_N=\Omega\left(N^{\epsilon -1}\right)$}  \;,
\ena$
\end{center}
where $\Pi_{t}$ is a non-negative matrix, whose elements are $\pi_t(i,j)=\pi(i,j)E_{h}\left\{ e^{t h j} \right\}$. $\left(\Pi_{t}\right)^N$ denotes the $N^{th}$ power of the matrix $\Pi_{t}$.
Let $D_t$ be the following diagonal matrix:\vspace{0.2cm}
\begin{center}
$\bea{l}
 D_t=
\left[\begin{matrix}
1 & 0  \\
0 & E_{h}\left\{ e^{t h} \right\}
\end{matrix}\right] .\vspace{0.2cm}
\ena$
\end{center}
Then, $\Pi_{t}$ can be rewritten as:
\beq
\label{eq:Pi_t}
\bea{l}
 \Pi_t=\Pi\cdot D_t=
 \left[\begin{matrix}
\pi(0,0) & (1-\pi(0,0))E_{h}\left\{ e^{t h} \right\}  \vspace{0.2cm}\\
1-\pi(1,1) & \pi(1,1)E_{h}\left\{ e^{t h} \right\}
\end{matrix}\right]\;.
\ena
\eeq
By applying the Perron-Frobenius Theorem \cite{Dembo_Large_1998} we have:
\beq
\label{eq:PF_Lambda}
\bea{l}
\displaystyle\Lambda(t)=\lim_{N\rightarrow\infty}\frac{1}{N}\Lambda_N(Nt)=\log \rho\left(\Pi_{t}\right),
\ena
\eeq
where $\rho\left(\Pi_{t}\right)$ denotes the Perron-Frobenius eigenvalue of the matrix $\Pi_{t}$, and is given by:
\beq
\bea{l}
\displaystyle\rho\left(\Pi_{t}\right)=\frac{\beta E_{h}\left\{ e^{t h} \right\}+\alpha}{2}\vspace{0.2cm}\\\hspace{0.5cm}
+\sqrt{\left(\frac{\beta E_{h}\left\{ e^{t h} \right\}+\alpha}{2}\right)^2-E_{h}\left\{ e^{t h}\right\}\left(\alpha\beta-(1-\alpha)(1-\beta)\right)},
\ena
\eeq
where $\alpha\triangleq\pi(0,0)$, $\beta\triangleq\pi(1,1)$, and $E_{h}\left\{ e^{t h}\right\}$ is the moment generating function of the fading channel. Note that $\rho\left(\Pi_{t}\right)$ is the isolated root of the characteristic equation of the matrix $\Pi_{t}$, positive, finite and differentiable with respect to $t$ \cite{lancaster1969theory}. Therefore, we can apply the Gartner-Ellis Theorem and Theorem \ref{th:fading_error_exponent}.b holds. In fact, we only need to require that the moment generating function of the channel fading be finite to guarantee that Statement (b) in Theorem \ref{th:fading_error_exponent} holds. Then, the rate function is given by $I_i(x)=\sup_{t\in \mathbb{R}}{\left( xt-\Lambda(t)\right)}$ under $H_i$, where $\rho\left(\Pi_{t}\right)$ under hypothesis $H_i$ is evaluated with respect to $\Pi_{t}$ governed by $H_i$.
\vspace{0.2cm}

\section{Simulation Results}\label{sec:simulations}
In this section we provide numerical examples illustrating detection performance under the SEEMA algorithm. The simulations were implemented in Matlab. We simulated a network that contains $N$ sensors. We simulated various scenarios which are captured by the theoretical analysis, including i.i.d. observations, non-i.i.d. observations (where the correlated observations were examined by Markovian models), equal channel gains, and Rayleigh fading channel gains. Other simulation parameters are described under each scenario in what follows.

We start by examining the detection of a Gaussian signal, which appears for example in radar signals, communication signals, and radio astronomy signals \cite{kay1998fundamentals, leshem2001multichannel, jayaweera2007bayesian}. The signal follows a distribution $\theta_n\sim\mathcal{N}(0,\sigma_{\theta,n}^2)$ independently across sensors, where $n$ denotes the sensor index, ($n=1,2,...,N$).
A random observation at sensor $n$ can be written under $H_0$ and $H_1$ as:
\beq
\bea{l}
H_0 \; : \; x_n=v_n  \;\;,\;\;
H_1 \; : \; x_n=\theta_n+v_n
\;,
\ena
\eeq
where $v_n\sim\mathcal{N}(0, \sigma_v^2)$ is the additive Gaussian observation noise, where we set $\sigma_v^2=1$. The observation noise is assumed to be i.i.d. across sensors. The transmission region was set to $\Gamma\triangleq\Gamma_n=\{x: X_L<|x_n|<\infty\} \;, \forall n$, where $X_L$ was set such that the average number of transmissions under SEEMA equaled $0.2\cdot N$. Note that a similar censoring-type transmission region can be applied when handling the multi-dimensional case as well, as discussed in Section \ref{ssec:implementation}.

\begin{figure}[htbp]
\begin{center}
    \subfigure[Error probability as a function of the number of sensors.]{\scalebox{0.6}
    {
      \label{fig:fig1a}{\epsfig{file=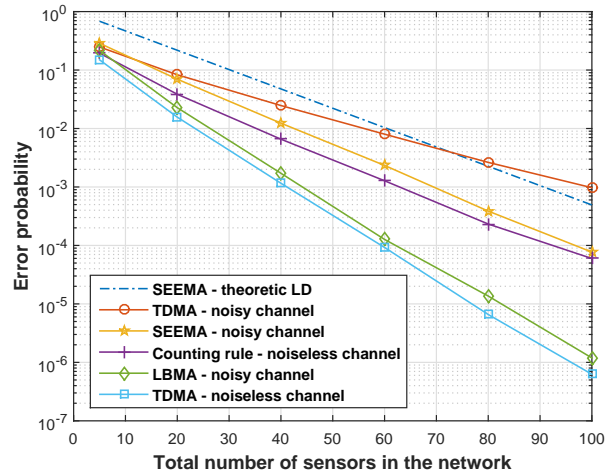}}
    }}
    \subfigure[Total average energy as a function of the error probability.]{\scalebox{0.6}
    {
      \label{fig:fig1b}{\epsfig{file=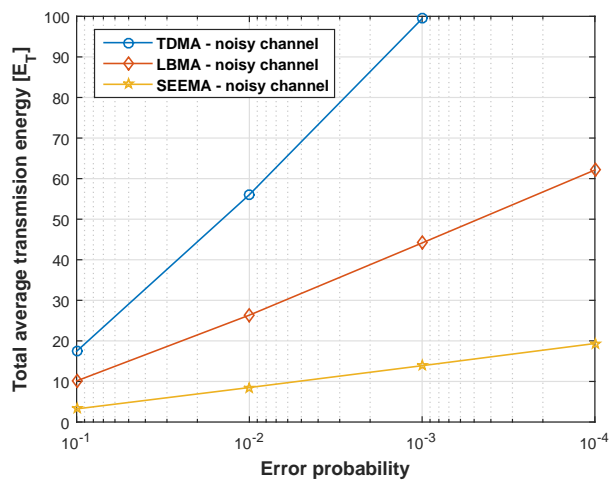}}
    }}
   \caption{Simulation parameters: Equal channel gain, i.i.d. observations under $H_i$, $E_N=E_T$ fixed.}
  \label{fig:fig1}
\end{center}
  \end{figure}

First, we examine the case of equal channel gains, as discussed in section \ref{ssec:equal}. We consider the case of i.i.d. observations under $H_i$, where we set $\sigma_{\theta,n}^2=3 , \forall n$. We obtained $X_L=1.9$. The channel AWGN was set to $w\sim\mathcal{N}(0,5)$. In addition to the SEEMA algorithm, we simulated the following algorithms for comparison: (i) the well-known TDMA scheme, where each sensor transmits its exact observation in a different time slot (i.e., using orthogonal noisy channels), referred to as the \emph{TDMA - noisy channel} in Fig. \ref{fig:fig1}. Note that similar to SEEMA, TDMA can be implemented with dumb sensors (oblivious to the observation statistics). Observation statistics are only needed at the FC. However, the bandwidth increases linearly with $N$ under the TDMA scheme, whereas only a single waveform is required under the SEEMA scheme. We simulated TDMA using a noise-free channel as well, to obtain a benchmark on detection performance, which is referred to as the \emph{TDMA - noiseless channel} in Fig. \ref{fig:fig1}. (ii) We have modified the counting rule in \cite{niu2006distributed} used for detecting deterministic signals in noise by first making local binary decisions at each sensor for random signals in noise. We have used the transmission region applied by SEEMA for making the local decisions, and a noiseless channel for transmission. Since we consider the equal channel gain case in this scenario, the received signal counts the local decisions (i.e., number of ones). Therefore, it serves as a benchmark for the detection performance under SEEMA, referred to as the \emph{Counting rule - noiseless cannel} in Fig. \ref{fig:fig1}. (iii) The LBMA scheme, where each sensor transmits its local LLR over a noisy MAC channel, is referred to as the \emph{LBMA - noisy channel} in Fig. \ref{fig:fig1}. Note that LBMA approaches optimal detection performance as the number of sensors increases since the noise term vanishes, and the summation of the LLRs in the independent observation case \cite{Liu_Type_2007}. However, LBMA requires knowing the observation distribution under each hypothesis at each sensor, and the hardware implementation is more complex than SEEMA since transmitting the random LLRs, which have a large dynamic range, can cause signal distortion from the saturation effect in the analog amplifiers. (iv) We also present the theoretical error probability (up to a constant factor), $e^{-nI}$, by computing the theoretical error exponent, $I$, in (\ref{eq:iid_no_fading_err_exponent}) proved by Theorem \ref{th:no_fading}.d.

Fig. \ref{fig:fig1} confirms the results of Theorem \ref{th:no_fading}. We set $E_N=E_T=1$ fixed. Fig. \ref{fig:fig1a} shows that the error probability decays exponentially with the total number of sensors in the network $N$, and achieved the theoretical error exponent (\ref{eq:iid_no_fading_err_exponent}) in Theorem \ref{th:no_fading}. Note that SEEMA outperformed TDMA in the noisy channel scenario, although all the sensors transmitted their exact measurements under the TDMA scheme. This is because TDMA suffers from channel noise in each dimension, which becomes negligible under the SEEMA scheme (due to the single-dimension transmission). In Fig. \ref{fig:fig1b}, we compare the performance of the TDMA, LBMA, and SEEMA algorithms for noisy channels in terms of average transmission energy as a function of the error probability. It can be seen that SEEMA significantly outperforms both TDMA and LBMA in terms of energy efficiency, thanks to its self censoring-type transmission scheme.

\begin{figure}[htbp]
\centering \epsfig{file=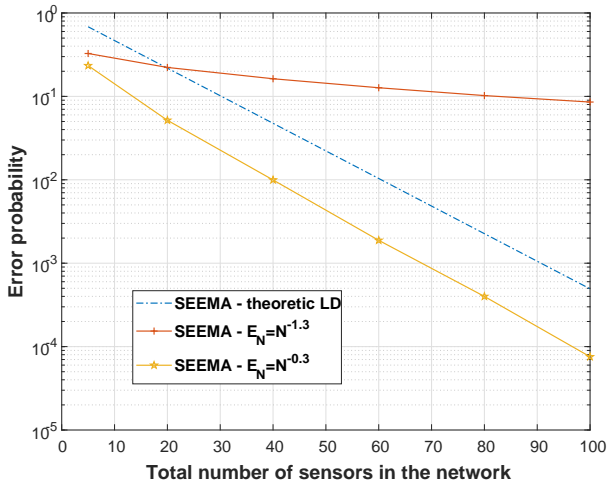,
width=0.5\textwidth}
\caption{Error probability as a function of the number of sensors. Simulation parameters: Equal channel gain, i.i.d. observations under $H_i$, transmission energy decreases with $N$.}
\label{fig:fig2}
\end{figure}

Fig. \ref{fig:fig2} confirms the results of Theorem \ref{th:no_fading} and Remark 2 regarding the performance of SEEMA. Here we set the channel AWGN to $w\sim\mathcal{N}(0,1)$. The error probability decayed exponentially with $N$ when $E_N=N^{-0.3}$ and achieved the theoretical error exponent (\ref{eq:iid_no_fading_err_exponent}) in Theorem \ref{th:no_fading}, but decayed sub-exponentially with $N$ when $E_N=N^{-1.3}$.

\begin{figure}[htbp]
\begin{center}
    \subfigure[Error probability as a function of the number of sensors.]{\scalebox{0.6}
    {
      \label{fig:fig3a}{\epsfig{file=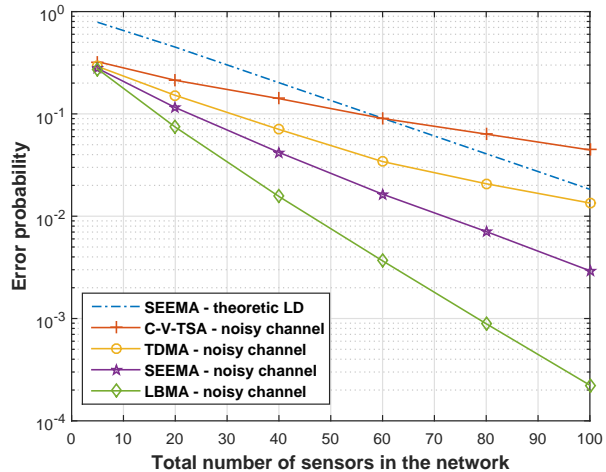}}
    }}
    \subfigure[Total average energy as a function of the error probability.]{\scalebox{0.6}
    {
      \label{fig:fig3b}{\epsfig{file=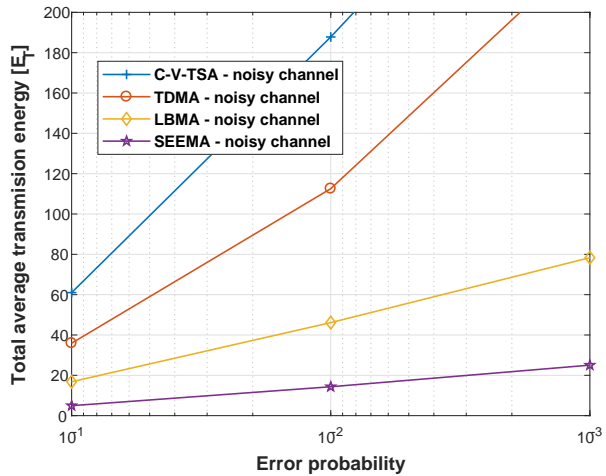}}
    }}
   \caption{Simulation parameters: i.i.d. Rayleigh fading channel, $h_n\sim \mbox{Rayleigh}(\sigma_h)$, i.ni.d. observations under Hi, $E_N=E_T$ fixed.}
  \label{fig:fig3}
\end{center}
  \end{figure}

Next, we examine the case where the sensors experience an i.i.d. Rayleigh fading channel gain, $h_n\sim \mbox{Rayleigh}(\sigma_h)$ (where the phase uncertainty is eliminated, as discussed in section \ref{ssec:fading}).
We considered the case of independent but non-identically distributed (i.ni.d.) observations under $H_i$. Specifically, we assumed that $N/2$ sensors observe $\theta_n\sim\mathcal{N}(0,3), n=1, ..., N/2$ and $N/2$ sensors observe $\theta_n\sim\mathcal{N}(0,4), n=N/2+1, ..., N$ (i.e., non-identically distributed observations). We set $E_N=E_T=1$ fixed. We obtained $X_L=1.97$.
In addition to SEEMA, we have simulated the following algorithms for comparison: (i) the TDMA scheme, where each sensor transmits its LLR in a different time slot (i.e., using orthogonal noisy fading channels), referred to as the \emph{TDMA - noisy channel}. (ii) The LBMA scheme, where each sensor transmits its local LLR over a noisy fading MAC channel, is referred to as the \emph{LBMA - noisy channel}. (iii) The Chair-Varshney fusion rule-based two-stage approximation (C-V-TSA) method that was investigated in [5]. The C-V-TSA method first makes local binary decisions at each sensor node, and makes a second binary decision for each signal received at the FC, which is multiplied by the channel gain. The detection statistics is then evaluated for decision. Note that the bandwidth increases linearly with $N$ under the C-V-TSA method. (iv) We also present the theoretical error probability (up to a constant factor), $e^{-nI}$, by computing the theoretical error exponent, $I_i(x)=\sup_{t\in \mathbb{R}}{\left( xt-\Lambda(t)\right)}$ under $H_i$, for detection threshold $x$, where $\Lambda(t)$ is computed by the closed-form expression developed in (\ref{eq:lambda_local_iid}).
Fig. \ref{fig:fig3} confirms the results of Theorem \ref{th:fading_error_exponent}. We set $E_N=E_T=1$ fixed. Fig. \ref{fig:fig3a} shows that the error probability decayed exponentially with the total number of sensors in the network $N$ and achieved the theoretical error exponent (\ref{eq:iid_no_fading_err_exponent}). Note that SEEMA outperformed TDMA in the noisy channel scenario again. Fig. \ref{fig:fig3b} compares the performance of C-V-TSA, TDMA, LBMA, and SEEMA algorithms in terms of average transmission energy as a function of the error probability. SEEMA significantly outperformed the other algorithms, as a result of its self censoring-type transmission scheme.

\begin{figure}[htbp]
\begin{center}
    \subfigure[Error probability as a function of the number of sensors.]{\scalebox{0.6}
    {
      \label{fig:fig4a}{\epsfig{file=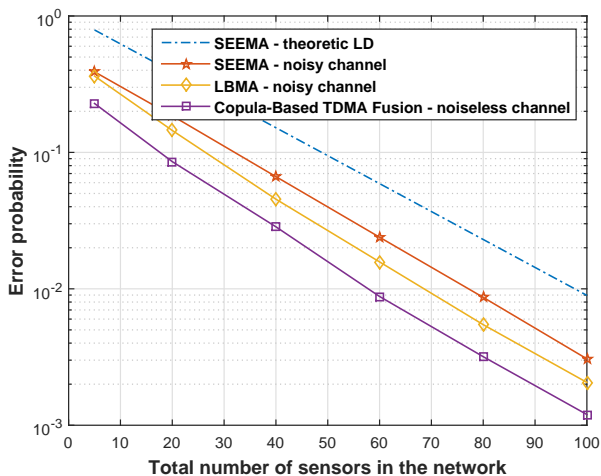}}
    }}
    \subfigure[Total average energy as a function of the error probability.]{\scalebox{0.6}
    {
      \label{fig:fig4b}{\epsfig{file=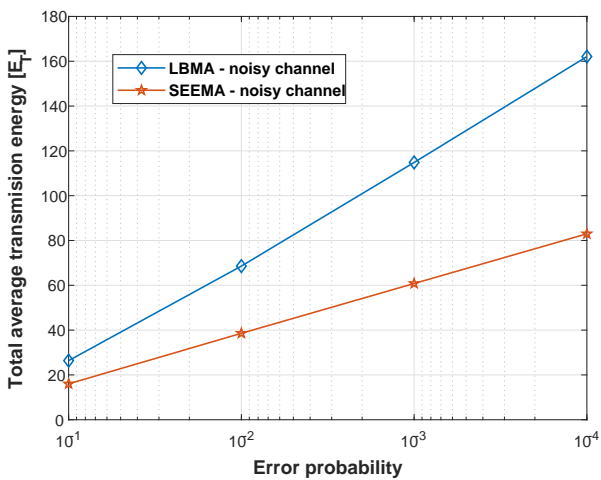}}
    }}
   \caption{Simulation parameters: Equal channel gain, correlated Markovian observations across sensors, $E_N=E_T$ fixed.}
  \label{fig:fig4}
\end{center}
  \end{figure}

Finally, we examine the case of correlated Markovian observations across sensors as analyzed in Section \ref{ssec:special_Markovian}. Specifically, each sensor measures an observation $0$ or $1$. Under hypothesis $H_0$, the transition probabilities were set to: $\pi_0(0,0)=0.65$, $\pi_0(1,1)=0.35$. Under hypothesis $H_1$, the transition probabilities were set to: $\pi_1(0,0)=0.35$, $\pi_1(1,1)=0.65$. We consider equal channel gains. In addition to SEEMA, we have simulated the following algorithms for comparison: (i) the TDMA scheme with copula-based fusion, where each sensor transmits its observation (one or zero) in a different time slot (i.e., using orthogonal noiseless channels), referred to as the \emph{Copula-Based TDMA Fusion - noiseless channel}. Then, the optimal detector computes the LLR forward based on the transition matrix for detection, which simplifies the implementation as compared to the general correlated case \cite{sundaresan2011copula}. Note that the bandwidth requirement increases linearly with $N$ under this scheme. (ii) The LBMA scheme, where each sensor transmits its local LLR over a noisy fading MAC channel, referred to as the \emph{LBMA - noisy channel}. (iii) We also present the theoretical error probability (up to a constant factor), $e^{-nI}$, by computing the theoretical error exponent, $I_i(x)=\sup_{t\in \mathbb{R}}{\left( xt-\Lambda(t)\right)}$ under $H_i$, for detection threshold $x$, where $\Lambda(t)$ is computed by the closed-form expression developed in (\ref{eq:PF_Lambda}).
Fig. \ref{fig:fig4} confirms the results of Theorem \ref{th:fading_error_exponent} and the analysis in Section \ref{ssec:special_Markovian}. We set $E_N=E_T=1$ fixed. Fig. \ref{fig:fig4a} shows that the error probability decays exponentially with the total number of sensors in the network $N$ and achieves the theoretical error exponent. In Fig. \ref{fig:fig4b}, we compare the performance of LBMA, and SEEMA algorithms in terms of average transmission energy as a function of the error probability. SEEMA significantly outperforms LBMA in terms of energy-efficiency due to its self censoring-type transmission scheme.
\section{Conclusion}\label{sec:conclusion}
We proposed a Spectrum and Energy Efficient Multiple Access (SEEMA) scheme for spectrum and energy efficient detection in WSNs. In SEEMA, only sensors with highly informative observations transmit their data in each data collection using a common analog waveform.
SEEMA has important advantages for detection tasks in WSN. It is highly energy and bandwidth efficient as compared to existing methods because of its transmissions savings and narrowband transmission over MAC. It can be implemented by simple dumb sensors which simplifies the implementation for detection tasks in WSNs.
Both finite sample analysis and asymptotic analysis of the error probability have been established with respect to the network size, and conditions for obtaining exponential decay of the error were developed. Specific performance analysis has been developed for common non-i.i.d. observation scenarios, including local i.i.d. observations, and Markovian correlated observations. Numerical examples demonstrated the strong performance of SEEMA.

\bibliographystyle{ieeetr}

\end{document}